\documentclass[preprint,english,aps,prx,showpacs,superscriptaddress,preprintnumbers]{revtex4-1}

\usepackage{float}
\usepackage{amsmath}
\usepackage{amssymb}
\usepackage{bm}
\usepackage{braket}
\usepackage{graphicx}
\usepackage[usenames, dvipsnames, svgnames, table]{xcolor}

\usepackage{soul}

\newcommand{\ee}{\text{e}}
\newcommand{\ii}{\text{i}}
\renewcommand{\vec}[1]{\mathbf{#1}}
\renewcommand{\alt}[1]{\mathsf{#1}}
\newcommand{\altvec}[1]{\bm{\mathcal{#1}}}

\begin{document}

\title{Second Quantization of Scattering Modes of Absorptive Photonic Nanostructures}
\author{J. Oppermann}
\email{jens.oppermann@kit.edu}
\affiliation{Institute of Theoretical Solid State Physics, Karlsruhe Institute of Technology, 76131 Karlsruhe, Germany}
\author{J. Straubel}
\affiliation{Institute of Theoretical Solid State Physics, Karlsruhe Institute of Technology, 76131 Karlsruhe, Germany}
\author{K. S{\l}owik}
\affiliation{Institute of Physics, Nicolaus Copernicus University, 87-100 Toru{\'n}, Poland}
\author{C. Rockstuhl}
\affiliation{Institute of Theoretical Solid State Physics, Karlsruhe Institute of Technology, 76131 Karlsruhe, Germany}
\affiliation{Institute of Nanotechnology, Karlsruhe Institute of Technology, 76131 Karlsruhe, Germany}

\begin{abstract}
We present a quantization scheme for optical systems with absorptive losses, based on an expansion in the complete set of scattering solutions to Maxwell's equations. The natural emergence of both absorptive loss and fluctuations without introducing a thermal bath is demonstrated. Our model predicts mechanisms of absorption induced squeezing and dispersion mediated photon conversion.
\end{abstract}

\maketitle

\section{Introduction}
Optical structures on the micro- and nanometer scale have recently gained interest in the context of future quantum technologies, as they can be used for the generation of quantum states of light \cite{Moreau2001, McKeever2004, Filter2014, Straubel2016} and coherent optical control of quantum states \cite{Mabuchi2002, Verhagen2012}. As the possibility of practical applications grows ever larger on the horizon, the need for efficient theoretical schemes to describe such systems quantum mechanicaly grows with them. Unfortunately, since the quantum theory of light-matter interaction started with the study of closed microwave cavities, most quantum models of light assume the existence of a discrete set of eigenmodes whereas optical systems are inherently open and, therefore, feature an uncountable number of scattering modes. Scattering modes are not eigenmodes in the usual sense, since they contain an incident field, which excites the system, as well as a scattered field containing only outgoing contributions. The quantitative quantum description of optical systems, therefore, often requires a large amount of heuristic modeling, phenomenological inference and/or experimental input. In two earlier contributions we tried to overcome this problem by first achieving proper field normalization of scattering modes \cite{Oppermann2018b} as a necessary step towards quantizing the field and then demonstrating how discrete eigenmodes can emerge from the resulting quantum field theory in systems with optical resonances \cite{Oppermann2018a}. In the process we found means to determine light-matter coupling paramters quantitatively, discussed the emergence of dissipative dynamics due to radiation losses and established an input-output formalism linking cavity and scattering dynamics. However, we did not account for absorptive losses due to a non-vanishing imaginary part of the permittivity. This is because the usual quantum field theory of light contains a large number of uncoupled and, therefore, freely evolving harmonic oscillator modes, while the presence of losses must introduce additional dynamics due to the fluctuation-dissipation theorem.

One commonly adapted approach to second quantization in absorptive systems was developed by Gruner and Welsch in their seminal work \cite{Gruner1996}. It consists of introducing quantum fluctuations of the electric charge and current densities, as warranted by the fluctuation-dissipation theorem, formally solving Maxwell's equations in the presence of said source fluctuations, and finally replacing the charge and current noise with its associated quantum operator, leading to a source operator expansion of the electric field. This technique has been succesfully employed in a number of applications in light-matter interaction \cite{Dung2002, VanVlack2012, Martin-Cano2014, Martin-Cano2016}. Notably, a recent contribution by Franke \textit{et al.} \cite{Franke2018} used the same concept of noise operators to perform the second quantization of quasinormal modes. In contrast to scattering modes, quasinormal modes form a discrete set of eigenmodes for open optical systems, but they exhibit a number of undesirable features such as energy non-conservation and exponential growth towards infinity \cite{Lai1990, Leung1994, Ching1996, Kristensen2012, Sauvan2013, Bai2013, Kristensen2013, Ge2014, Muljarov2016}. A thorough comparison between scattering modes and quasinormal modes can be found in App. C of \cite{Oppermann2018b}.

The above approach lies in contrast to the canonical quantization scheme for photons in free space \cite{Vogel2006}, where quantization is performed with solutions to Maxwell's equations in the absence of charges. This shows that the problem of field quantization can be approached from several different angles, with each one yielding different physical insights. For this reason, we present here an alternative approach to second quantization of absorptive systems based on normalized scattering solutions to Maxwell's equations. The scientific value of this approach is validated by the occurence of additional terms and effects not found when following the procedure according to Gruner and Welsch \cite{Gruner1996}. We demonstrate how a series of simple modifications to the process of second quantization in free space allows us to find a quantum field theory of light in the presence of arbitrary but finite optical structures, where radiative and absorptive losses are encoded in the classical part of the field operators. Although the resulting Hamiltonian remains fully unitary, losses manifest under a change of basis as discussed in \cite{Oppermann2018a}. Compared to the lossless case, the system Hamiltonian exhibits additional terms that give rise to more complex dynamics. One consequence of this is the emergence of a photon background in accordance with the fluctuation-dissipation theorem, which emerges naturally instead of being assumed at the beginning. Furthermore, we predict mechanisms of intrinsic light squeezing in absorptive optical cavities and dispersion mediated photon conversion.

This work is organized as follows. In Sec. \ref{sec:quantization} we perform the second quantization of absorptive optical systems on a semiclassical basis, i.e. by using the complex permittivity function $\epsilon(\omega, \vec{x})$ to describe bulk matter. The resulting Hamiltonian is Hermitian but does not conserve photon number. In Sec. \ref{sec:cQED} we show that absorptive losses are present in the model, neglecting all but the free photon energy terms in the Hamiltonian. Absorption is encoded in the classical part of the electric field operator and can lead to a change in photon number under a transformation from input to output states. In Sec. \ref{sec:squeezing} the influence of counter-rotating terms in the Hamiltonian on the system dynamics and photon statistics is discussed. We find that the counterrotating part of the Hamiltonian modifies the ground state to describe a bunched photon background and can lead to ground state squeezing of Lorentzian cavity modes. Section \ref{sec:coupling} discusses dispersive photon conversion terms in the Hamiltonian. We show that these terms do not lead to modifications of classical scattering amplitudes, but can give rise to off-resonant coupling between cavity modes.

\section{Quantization Procedure}\label{sec:quantization}
We consider a localized optical system described by a spacially inhomogeneous and complex permittivity function $\epsilon(\omega, \vec{x})$, which fits within a virtual sphere of radius $R$ around the origin. Outside of the virtual sphere we assume a homogeneous and lossless medium, i.e. $\epsilon(\omega, \left|\vec{x}\right| > R) = \epsilon_b \in \mathbb{R}$. This last condition guarantees the existence of a set of normalizable and orthogonal scattering modes \cite{Oppermann2018b}. In the Coulomb gauge and in the absence of external charges, the classical Lagrangian of such a system can be written as
\begin{align}
\mathcal{L} =& \frac{\epsilon_0}{2}\dot{\vec{A}}\cdot\dot{\vec{C}} - \frac{1}{2\mu_0} \left(\vec{\nabla} \times \vec{A}\right)^2 + \vec{A}\cdot\vec{j},
\end{align}
where $\vec{A}$ and $\vec{C}$ are the vector potentials of the electric field $\vec{E} = -\partial_t \vec{A}$ and electric displacement $\vec{D} = -\epsilon_0\partial_t \vec{C}$ and $\vec{j}$ is the internal electric current induced in the system. Please note that we do not absorb the induced current into the electric displacement, as is usually done. We do this to separate the Lagrangian into absorptionless and absorptive parts.

The classical Hamiltonian of the system reads
\begin{align}
\label{eq:classical_hamiltonian}
\mathcal{H} &= 2\epsilon_0^{-1}\vec{\Pi}_A\vec{\Pi}_C + \frac{1}{2\mu_0} \left(\vec{\nabla}\times\vec{A}\right)^2 - \vec{A} \cdot \vec{j},
\end{align}
where $\vec{\Pi}_A$ and $\vec{\Pi}_B$ are the conjugate momenta to $\vec{A}$ and $\vec{C}$ reading
\begin{align}
\label{eq:canonical_momentum_a}
\vec{\Pi}_A =& \frac{\partial\mathcal{L}}{\partial\dot{\vec{A}}} = \frac{\epsilon_0}{2}\dot{\vec{C}}\\
\label{eq:canonical_momentum_c}
\vec{\Pi}_C =& \frac{\partial\mathcal{L}}{\partial\dot{\vec{C}}} = \frac{\epsilon_0}{2}\dot{\vec{A}}
\end{align}
To proceed with the second quantization of the electromagnetic field, we first redefine the classical fields in Eq. \eqref{eq:classical_hamiltonian} as operators and then rearrange them to enforce hermiticity. The resulting Hamiltonian density operator reads
\begin{align}
\label{eq:quantized_hamiltonian}
\hat{\mathcal{H}} &= \epsilon_0^{-1}\hat{\vec{\Pi}}_A\hat{\vec{\Pi}}_C + \epsilon_0^{-1}\hat{\vec{\Pi}}_C\hat{\vec{\Pi}}_A + \frac{1}{2\mu_0} \left(\vec{\nabla}\times\hat{\vec{A}}\right)^2-\frac{1}{2}\left(\hat{\vec{A}}\hat{\vec{j}}+\hat{\vec{j}}\hat{\vec{A}}\right),
\end{align}
where the notation $\hat{\cdot}$ for operators is used for clarity, but will be dropped from this point on. Since $\vec{A}$ and $\vec{\Pi}_A$ are conjugate variables, their commutator has to be of the canonical form \cite{Vogel2006}
\begin{align}
\label{eq:canonical_commutation_relation}
\left[\vec{A}(\vec{x}), \vec{\Pi}_A(\vec{x'})\right] =& \ii\frac{\hbar}{2}\delta^{(3)}(\vec{x}-\vec{x'}).
\end{align}
Please note that the factor $1/2$ in Eq. \eqref{eq:canonical_commutation_relation} is required to include second quantization in free space as a special case. To be more specific, in free space $\vec{C} = \vec{A}$ and the factor $1/2$ in Eq. \eqref{eq:canonical_momentum_a} is not present. To correct for this, Eq. \eqref{eq:canonical_commutation_relation} has to be multiplied by $2$, leading to the well-known free space commutator.

The next step is to expand the field operators into a basis of solutions to Maxwell's wave equation. We do this in frequency space, seeking time harmonic solutions for the electric field whose spatial dependency obeys
\begin{align} \label{eq:wave_equation}
\vec{\nabla}\times\vec{\nabla}\times\vec{E}(\omega, \vec{x}) - \epsilon(\omega, \vec{x})\frac{\omega^2}{c_0^2}\vec{E}(\omega, \vec{x}) =& 0.
\end{align}
The set of wave solutions we chose are scattering modes of the form
\begin{align}
\label{eq:scattering_modes}
\vec{E}_{\vec{k}, \lambda} =& \left[\frac{\hbar\omega}{(2\pi)^32\epsilon_0\epsilon_b}\right]^{1/2} \hat{\vec{e}}_{\vec{k}, \lambda} \ee^{\ii\vec{k}\vec{x}} + \vec{E}^{(\vec{k})}_{\text{sca}}(\vec{x}),
\end{align}
where $\vec{k}$ and $\lambda=1,2$ label the wave vector and polarization of the incident plane wave, $\epsilon_b$ denotes the permittivity of the absorptionless environment, $\hat{\vec{e}}_{\vec{k}, \lambda}$ is a unit polarization vector and the scattered field $\vec{E}^{(\vec{k})}_{\text{sca}}(\vec{x})$ is defined to contain only outgoing contributions, i.e. it satisfies Silver-M\"uller radiation conditions. In \cite{Oppermann2018b} it is demonstrated that the scattering modes are complete, orthogonal and normalized with respect to the scalar product
\begin{align}
\int \text{d}^3x \epsilon_0\frac{\epsilon'(\omega, \vec{x})+\epsilon'(\omega', \vec{x})}{2} \vec{E}^*_{\vec{k}, \lambda}(\vec{x})\cdot\vec{E}_{\vec{k'}, \lambda'}(\vec{x}) =& \frac{\hbar\omega}{2} \delta(\vec{k}-\vec{k'})\delta_{\lambda,\lambda'}.
\end{align}
Please note that we choose here a symmetrized form of the scalar product without the frequency derivative of the permittivity, as opposed to the form used in \cite{Oppermann2018b}. This choice does not influence the correct normalization of the scattering modes, since the normalization only depends on the far field and the background medium is necessarily non-absorptive and non-dispersive for a system supporting scattering modes. The reason for choosing this specific form of scalar product is that we directly encounter it during the second quantization procedure shown here. If one were to perform second quantization in terms of the finite-bandwidth, finite-energy (physical) modes introduced in \cite{Oppermann2018b} and only perform the plane wave (zero bandwidth, infinite energy) limit afterwards, one would encounter the dispersive form of the normalization integral. In the end, of course, the resulting Hamiltonian is the same whether one performs the plane wave limit before or after quantization. 

We choose the following expansion of the operators $\vec{A}$ and $\vec{\Pi}_A$
\begin{align}
\label{eq:a_expansion}
\vec{A}(\vec{x}) =& \sum_{\lambda=1,2}\int \text{d}^3k \omega^{-1} \left[\vec{E}_{\vec{k},\lambda}(\vec{x})a_{\vec{k},\lambda} + \vec{E}^*_{\vec{k},\lambda}(\vec{x})a^\dagger_{\vec{k},\lambda}\right] \\
\label{eq:pia_expansion}
\vec{\Pi}_A(\vec{x}) =& \sum_{\lambda=1,2}\int \text{d}^3k \frac{1}{2\ii}\epsilon_0\epsilon'(\omega, \vec{x}) \left[\vec{E}_{\vec{k},\lambda}(\vec{x})a_{\vec{k},\lambda} - \vec{E}^*_{\vec{k},\lambda}(\vec{x})a^\dagger_{\vec{k},\lambda}\right],
\end{align}
with the creation and annihilation operators satisfying
\begin{align}
\label{eq:scattering_mode_commutator}
\left[a_{\vec{k},\lambda}, a^\dagger_{\vec{k'},\lambda'}\right] = \delta(\vec{k}-\vec{k'})\delta_{\lambda,\lambda'},
\end{align}
and all other commutators vanishing. Using Eqs. \eqref{eq:a_expansion} through \eqref{eq:scattering_mode_commutator} it is easy to verify that the canonical commutation relation \eqref{eq:canonical_commutation_relation} is fulfilled and the chosen expansion is therefore valid.

In order to complete the quantization, we still need to find operator expansions of $\vec{\Pi}_C$ and $\vec{j}(\vec{x})$. To this end, we make the assumption that the classical constitutive relations continue to hold in the operator description. They are given in terms of the complex permittivity $\epsilon(\omega, \vec{x})$ function in frequency domain according to
\begin{align}
\label{eq:constitutive_relation_d}
\vec{D}(\omega, \vec{x}) &= \epsilon_0\epsilon'(\omega, \vec{x})\vec{E}(\omega, \vec{x}), \\
\label{eq:constitutive_relation_j}
\vec{j}(\omega, \vec{x}) &= \omega\epsilon_0\epsilon''(\omega, \vec{x})\vec{E}(\omega, \vec{x}),
\end{align}
where $\epsilon'$ and $\epsilon''$ denote the real and imaginary part of the permittivity, respectively. We identify
\begin{align}\label{eq:e_to_pi}
\vec{E}(\vec{x}) =& -2\epsilon^{-1}\vec{\Pi}_C(\vec{x}), \\
\label{eq:d_to_pi}
\vec{D}(\vec{x}) =& -2\vec{\Pi}_A(\vec{x}),
\end{align}
according to Eqs. \eqref{eq:canonical_momentum_a} and \eqref{eq:canonical_momentum_c}. Our goal is to find expansions of $\vec{\Pi}_C$ and $\vec{j}$ in photon creation and annihilations operators, which satisfy Eqs. \eqref{eq:e_to_pi} and \eqref{eq:d_to_pi} as well as the constitutive relations in Eqs. \eqref{eq:constitutive_relation_d} and \eqref{eq:constitutive_relation_j}. This leads to the unique choice of expansions
\begin{align} \label{eq:pic_expansion}
\vec{\Pi}_C(\vec{x}) =& \sum_{\lambda=1,2}\int \text{d}^3k \frac{1}{2\ii}\epsilon_0\left[\vec{E}_{\vec{k},\lambda}(\vec{x}) a_{\vec{k},\lambda} - \vec{E}^*_{\vec{k},\lambda}(\vec{x}) a_{\vec{k},\lambda}^\dagger\right] \\
\label{eq:j_expansion}
\vec{j}(\vec{x}) =& \sum_{\lambda=1,2}\int \text{d}^3k \ii\omega\epsilon_0\epsilon''(\omega, \vec{x})\left[\vec{E}_{\vec{k},\lambda}(\vec{x}) a_{\vec{k},\lambda} - \vec{E}^*_{\vec{k},\lambda}(\vec{x}) a_{\vec{k},\lambda}^\dagger\right]
\end{align}

All that now remains to be done is inserting the operator expansions \eqref{eq:a_expansion}, \eqref{eq:pia_expansion}, \eqref{eq:pic_expansion} and \eqref{eq:j_expansion} into the system Hamiltonian \eqref{eq:quantized_hamiltonian} and integrating over space. In order to eliminate the curl operators we can use that the expansion fields satisfy the wave equation \eqref{eq:wave_equation}. The calculations are straightforward and presented in App. \ref{app:hamiltonian_derivation}. The resulting Hamiltonian reads
\begin{align} \label{eq:complete_hamiltonian}
H =& \sum_{\lambda=1,2}\int \text{d}^3k \hbar\omega a^\dagger_{\vec{k},\lambda}a_{\vec{k},\lambda} \nonumber \\
&- \sum_{\lambda,\lambda'} \int \text{d}^3k \text{d}^3k' \left[\zeta(\vec{k}, \lambda; \vec{k'}, \lambda') a_{\vec{k},\lambda}a_{\vec{k'},\lambda'} + \zeta^*(\vec{k}, \lambda; \vec{k'}, \lambda') a^\dagger_{\vec{k},\lambda}a^\dagger_{\vec{k'},\lambda'}\right] \nonumber \\
&- \sum_{\lambda,\lambda'} \int \text{d}^3k \text{d}^3k' \kappa(\vec{k}, \lambda; \vec{k'}, \lambda') a^\dagger_{\vec{k},\lambda}a_{\vec{k'},\lambda'},
\end{align}
where the functions $\zeta$ and $\kappa$ are defined as
\begin{align}\label{eq:thermal_coefficient}
\zeta(\vec{k}, \lambda; \vec{k'}, \lambda') = \int \text{d}^3x \frac{\ii}{4}\epsilon_0\left[\epsilon''(\omega, \vec{x})+\epsilon''(\omega', \vec{x})\right] \vec{E}_{\vec{k},\lambda}(\vec{x})\cdot\vec{E}_{\vec{k'},\lambda'}(\vec{x}), \\
\label{eq:coupling_coefficient}
\kappa(\vec{k}, \lambda; \vec{k'}, \lambda') = \int \text{d}^3x \frac{\ii}{2}\epsilon_0\left[\epsilon''(\omega, \vec{x})-\epsilon''(\omega', \vec{x})\right] \vec{E}^*_{\vec{k},\lambda}(\vec{x})\cdot\vec{E}_{\vec{k'},\lambda'}(\vec{x}).
\end{align}
Please note that the functions $\zeta$ and $\kappa$ are calculated as spatial overlap integrals of scattering solutions at different frequencies $\omega$. Therefore, spectral as well as spatial overlap of scattering solutions is of importance.

The above Hamiltonian consists of three contributions. The first one just describes harmonic photon phase oscillations and is the only one also present in the absence of absorptive losses. However, we will see that even in the absence of other terms absorption manifests in the input-output relations of the system, due to absorption being already encoded in the classcial part of the operator expansions. The second term consists of squeezing terms, which violate the conservation of photon number. Equation \eqref{eq:thermal_coefficient} shows that this term arises due to the presence of absorptive losses. The third term describe coupling between different scattering modes. According to Eq. \eqref{eq:coupling_coefficient} this term arises due to dispersion. Please note that in realistic materials dispersion and absorption always occur hand-in-hand according to the Kramers-Kronig relations.

The remainder of this work is devoted to two goals. First, we need to determine whether the Hamiltonian in Eq. \eqref{eq:complete_hamiltonian} can reproduce the phenomenology of systems with absorptive losses, especially the loss of photons. Second, we want to search for any new light-matter phenomena predicted by the model. In the following sections the individual terms of Eq. \eqref{eq:complete_hamiltonian} are studied one at a time, both in the context of scattering theory and in the context of cavity quantum electrodynamics (cQED). In Sec. \ref{sec:cQED} we show that the harmonic term together with the operator expansion of the electric field is already sufficient to describe photon absorption. The squeezing terms are discussed in detail in Sec. \ref{sec:squeezing}, where it is shown to produce a bunched photon background in arbitrary absorptive structures as well as squeezing in optical cavities. Finally, in Sec. \ref{sec:coupling} we show that the photon conversion term does not lead to a modification of a structures optical spectrum, but can lead to dispersive cavity-cavity coupling in cQED scenarios.

\section{Harmonic Term} \label{sec:cQED}
Since the free photon energy term is unmodified compared the lossless case, one might not expect it to describe any loss-related phenomena. However, the photon annihilation and creation operators gain their physical meaning only in the context of the field expansion in Eq. \eqref{eq:a_expansion}. Since the classical eigenmodes of the system are already subjected to losses, absorptive losses can be introduced into the freely evolving quantum system. The key to finding dissipation in a quantum context is the input-output formalism introduced in \cite{Oppermann2018a}. In this section, we demonstrate how photon absorption can be described in scattering and cQED scenarios by expanding the field operator in appropriate input and output bases.

\subsection{Scattering Theory}
In terms of the input-output formalism, the central difference between absorptive and non-absorptive systems is the construction of output states. For non-absorptive systems, output states can be calculated from input states by applying time-reversal \cite{Oppermann2018a}. In absorptive systems this procedure fails, because the wave equation \eqref{eq:wave_equation} is no longer invariant under time-reversal. Therefore, an asymmetry between input and output modes is naturally introduced due to them arising from different equations. In the following, we show that this asymmetry leads to photon loss.

Let us consider a scattering scenario with absorptive losses. In App. \ref{app:input_output_relation} it is shown that the input and output states are connected according to
\begin{align} \label{eq:input_output_relation}
a_{\vec{k},\lambda}^{(\text{out})} = a_{\vec{k},\lambda}^{(\text{in})} + \frac{\ii}{2\pi} \sum_{\lambda'=1,2} \int_{k'=k} d\vec{\hat{k}'} \frac{\altvec{A}^{(\vec{k'},\lambda')}(\vec{\hat{k}})\cdot\vec{E}_0^{(\vec{k},\lambda)}}{\left|\vec{E}_0^{(\vec{k},\lambda)}\right|^2} a_{\vec{k'},\lambda'},
\end{align}
where $\altvec{A}^{(\vec{k'})}(\vec{\hat{k}})$ is the scattering amplitude of an incident photon with momentum $\vec{k'}$ into the direction $\vec{\hat{k}}$ and $\vec{E}_0^{(\vec{k},\lambda)}$ is the single-photon field amplitude
\begin{align}
\vec{E}_0^{(\vec{k},\lambda)} = \left[\frac{\hbar\omega}{(2\pi)^32\epsilon_0\epsilon_b}\right]^{1/2} \hat{\vec{e}}_{\vec{k}, \lambda}.
\end{align}
Please note that the output mode operators obey bosonic commutation relations in the physical plane wave limit discussed in \cite{Oppermann2018b}.

Assume now that a single photon-state of finite but large spacial extent is incident on the system, described by the state vector
\begin{align} \label{eq:beam_state}
\ket{\Phi} =& \int \frac{\text{d}^3k}{k} f(\hat{\vec{k}}) \sqrt{\frac{\Delta k}{2\pi}} \frac{1}{k-k_0-\ii \Delta k/2} a^{(\text{in})\dagger}_{\vec{k},\lambda_0}, \\
f(\hat{\vec{k}}) =& \frac{1}{\sqrt{2\pi}}\left(\sqrt{\frac{2}{\pi}}\delta^{-1}\right)^{1/2} \exp\left(-\left[\frac{\cos(\Theta) - 1}{2\delta}\right]^2\right), \nonumber
\end{align}
where $\Delta k$ and $\delta$ are the longitudinal and transversal momentum spread, respectively, which are inversely proportional to the spacial length and width of the pulse. It is easily verified that
\begin{align}
\braket{\Phi|\Phi} =& 1, \\
\sum_{\lambda=1,2}\int \text{d}^3k \braket{\Phi|a^{(\text{in})\dagger}_{\vec{k},\lambda}a^{(\text{in})}_{\vec{k},\lambda}|\Phi} =& 1.
\end{align}
Assuming that $\Delta k$ and $\delta$ are small compared to the typical scale of variations in $\altvec{A}^{(\vec{k'})}(\vec{\hat{k}})$ and using Eq. \eqref{eq:input_output_relation}, it is straightforward to calculate
\begin{align} \label{eq:output_photon_number_beam}
\sum_\lambda \int \text{d}^3k \braket{a^{(\text{out})\dagger}_{\vec{k}, \lambda}a^{(\text{out})}_{\vec{k}, \lambda}} \approx 1-\frac{\delta}{2\pi}&\left\{4\pi\text{Im}\left[\frac{\altvec{A}^{(\vec{k}_0,\lambda_0)}(\vec{\hat{k}_0})\cdot\vec{E}_0^{(\vec{k}_0,\lambda_0)}}{\left|\vec{E}_0^{(\vec{k}_0,\lambda_0)}\right|^2}\right]\right. \nonumber \\
& \left.-\int_{k = k_0} \frac{\left|\altvec{A}^{(\vec{k}_0,\lambda_0)}(\vec{\hat{k}})\right|^2}{\left|\vec{E}_0^{(\vec{k}_0,\lambda_0)}\right|^2}\right\}.
\end{align}
The terms in curly brackets can be readily identified as the imaginary part of the forward scattering amplitude and the integrated scattering intensity. From scattering theory it is well known that these quantities are related to the extinction and scattering cross sections according to \cite{Novotny2012}
\begin{align} \label{eq:total_cross_section}
\sigma_{\text{tot}} =& \frac{4\pi}{k_0^2}\text{Im}\left[\frac{\altvec{A}^{(\vec{k}_0,\lambda_0)}(\vec{\hat{k}_0})\cdot\vec{E}_0^{(\vec{k}_0,\lambda_0)}}{\left|\vec{E}_0^{(\vec{k}_0,\lambda_0)}\right|^2}\right], \\
\label{eq:scattering_cross_section}
\sigma_{\text{sca}} =& \frac{1}{k_0^2}\int_{k = k_0} \frac{\left|\altvec{A}^{(\vec{k}_0,\lambda_0)}(\vec{\hat{k}})\right|^2}{\left|\vec{E}_0^{(\vec{k}_0,\lambda_0)}\right|^2}.
\end{align}
From beam optics it is known that the angular spread $\alpha$ and the waist radius $w$ of a Gaussian beam are related according to
\begin{align} \label{eq:beam_radius}
w =& \frac{\lambda}{\pi\alpha}.
\end{align}
According to the definition of $f(\vec{\hat{k}})$ in Eq. \eqref{eq:beam_state}, $\alpha$ is related to the spread parameter $\delta << 1$ according to
\begin{align} \label{eq:beam_angle}
\cos(\alpha) - 1 =& \delta~\Rightarrow~\alpha \approx \sqrt{2\delta}.
\end{align}
Combining Eqs. \eqref{eq:output_photon_number_beam} through \eqref{eq:beam_angle}, we find for the expected number of output photons
\begin{align} \label{eq:output_photon_number_beam_final}
\sum_\lambda \int \text{d}^3k \braket{a^{(\text{out})\dagger}_{\vec{k}, \lambda}a^{(\text{out})}_{\vec{k}, \lambda}} \approx& 1 - \sqrt{2\pi} \frac{\sigma_{\text{tot}}-\sigma_{\text{sca}}}{A} = 1-\sqrt{2\pi} \frac{\sigma_{\text{abs}}}{A},
\end{align}
where $A$ is the area of the beam. Equation \eqref{eq:output_photon_number_beam_final} tells us that the probability of absorbing a photon is proportional to the absorption cross section over the beam area, just as expected.

\subsection{Cavity Quantum Electrodynamics}
Assume that an atom is coupled to a leaky cavity mode of Gaussian lineshape. In \cite{Oppermann2018a} it is shown that the light-matter coupling strength for such a scenario reads
\begin{align} \label{eq:in_out_coupling_constant}
\alt{g}^{(\text{in/out})} &= \sqrt{\frac{G}{4\pi}} \frac{\pi}{c_0^{3/2}}\omega_0\sqrt{2\Gamma}\frac{\vec{E}^{(\text{in/out})}_{\text{max}}\cdot\vec{d}}{\hbar},
\end{align}
where $\omega_0$ and $\Gamma$ are the central frequency and linewidth of the cavity mode, $G$ is a parameter describing the angular aparture of the cavity, $\vec{d}$ is the transition dipole moment and $\vec{E}^{(\text{in/out})}_{\text{max}}$ is the field strength at the position of the emitter at resonance. Please note that in Eq. \eqref{eq:in_out_coupling_constant} we distinguish between an input and an output coupling strength, based on whether input or output modes are used in the calculation. This is not necessary in the absence of absorption losses, where time-reversal invariance implies $\alt{g}^{(\text{out})} = \alt{g}^{(\text{in})}$. Assume now that the light-matter coupling is weak and that the emitter is initially in its excited state. According to \cite{Oppermann2018a}, the probability of finding the emitter in its excited state at time $t$ is
\begin{align} \label{eq:spontaneous_emission_atom}
P_e(t) = \exp\left(-2\frac{\left|\alt{g}^{(\text{in})}\right|^2}{\Gamma/2}t\right).
\end{align}
Please note that $\Gamma$ is the total decay rate consisting of radiative and absorptive losses and that Eq. \eqref{eq:spontaneous_emission_atom} therefore already describes decreased lifetimes due to absorption. The question is, of course, whether the expected number of emitted photons is smaller than $1$. Following the calculations presented in \cite{Oppermann2018a}, the photon number at time $t$ can be easily calculated as
\begin{align} \label{eq:spontaneous_emission_photon}
\int \text{d}^3k \braket{a^{\dagger(\text{out})}_{\vec{k}, \lambda}a^{(\text{out})}_{\vec{k}, \lambda}}_t \approx \frac{\left|\alt{g}^{(\text{out})}\right|^2}{\left|\alt{g}^{(\text{in})}\right|^2} \left(1-P_e(t)\right).
\end{align}
Taking the limit $t \rightarrow \infty$, Eq. \eqref{eq:spontaneous_emission_photon} tells us that the number of emitted photons $N_p$ is just
\begin{align} \label{eq:spontaneous_emission_photon_number}
N_p = \frac{\left|\alt{g}^{(\text{out})}\right|^2}{\left|\alt{g}^{(\text{in})}\right|^2}.
\end{align}
For an absorptive system we, therefore, expect the outcoupling strength to be smaller than the incoupling strength. For systems with gain the opposite should hold.

The outcoupling strength can be easily calculated using the input-output relations derived in App. \ref{app:input_output_relation}. First, we derive the Heisenberg equations of motion for the output modes. The result reads
\begin{align}
\dot{a}_{\vec{k}, \lambda}^{\text{(out)}} = -\ii\omega a_{\vec{k}, \lambda}^{\text{(out)}} - \ii g_{\vec{k}, \lambda}^{\text{out}} \sigma_-,
\end{align}
where $\sigma_-$ is the lowering operator of the emitter and the mode resolved coupling constants $g_{\vec{k}, \lambda}^{\text{out}}$ read
\begin{align} \label{eq:mode_resolved_outcoupling}
\alt{g}_{\vec{k}, \lambda}^{\text{out}} &= \alt{g}_{\vec{k}, \lambda}^{\text{in}} + \frac{\ii}{2\pi}\sum_{\lambda'}\int_{k=k'}d\hat{\vec{k'}} \frac{\altvec{A}^{(\vec{k'},\lambda)}(\hat{\vec{k}}) \cdot \vec{E}_0^{*(\vec{k}, \lambda)}}{\left|\vec{E}_0^{(\vec{k}, \lambda)}\right|^2} \alt{g}_{\vec{k'}, \lambda'}^{\text{in}},
\end{align}
where the incoupling coefficients $\alt{g}_{\vec{k}, \lambda}^{\text{in}}$ take on the form (see \cite{Oppermann2018a})
\begin{align}
\alt{g}_{\vec{k}, \lambda}^{\text{in}} :&= \frac{\vec{E}_{0}(\vec{x}_e)\cdot\vec{d}}{\hbar}\sqrt{\frac{\Gamma}{2\pi}}\frac{g(\hat{\vec{k}})}{\omega-\omega_0-\ii\Gamma/2} \delta_{\lambda, \lambda_0},
\end{align}
where $\vec{x}_e$ is the spacial position of the emitter, $g(\hat{\vec{k}})$ is a function on the unit sphere describing the coupling efficiency's dependence on the direction of incidence, and $\lambda_0$ is the polarization state coupling to the emitter in a coupled-uncoupled basis. The integrated outcoupling strength can be calculated from Eq. \eqref{eq:mode_resolved_outcoupling} according to
\begin{align} \label{eq:outcoupling}
\left|\alt{g}^{\text{out}}\right|^2 = \sum_\lambda\int \text{d}^3k \left|\alt{g}_{\vec{k}, \lambda}^{\text{out}}\right|^2.
\end{align}
After a straightforward calculation, Eqs. \eqref{eq:spontaneous_emission_photon_number}, \eqref{eq:mode_resolved_outcoupling}, and \eqref{eq:outcoupling} give for the number of emitted photons
\begin{align}
N_p = 1 - \frac{1}{2\pi} \int_{k=k_0}d\hat{\vec{k}}d\hat{\vec{k'}} \frac{g^*(\hat{\vec{k}})g(\hat{\vec{k'}})}{G} & \left[\ii \frac{\altvec{A}^{(\vec{k})*}(\hat{\vec{k'}})\cdot\vec{E}_0^{(\vec{k'})}}{\left|\vec{E}_0^{(\vec{k'})}\right|^2} - \ii \frac{\altvec{A}^{(\vec{k'})*}(\hat{\vec{k}})\cdot\vec{E}_0^{(\vec{k})}}{\left|\vec{E}_0^{(\vec{k})}\right|^2} \right. \nonumber \\
& \left. - \frac{1}{2\pi} \int \text{d}\hat{\vec{k''}} \frac{\left(\altvec{A}^{(\vec{k})*}(\hat{\vec{k''}})\cdot\vec{E}_0^{(\vec{k''})}\right)\left(\altvec{A}^{(\vec{k'})}(\hat{\vec{k''}})\cdot\vec{E}_0^{(\vec{k''})*}\right)}{\left|\vec{E}_0^{(\vec{k''})}\right|^4}\right].
\end{align}
The above expression is rather complicated, since it contains mutually interfering contributions from many directions. However, for the special case of highly directional light-matter coupling, i.e. for $g(\hat{\vec{k}})$ being strongly confined around a direction $\hat{\vec{e}}$, the optical theorem is once more recovered and we find
\begin{align}
N_p \approx 1-\frac{1}{4\pi^2}k_0^2\sigma_{\text{abs}}(\hat{\vec{e}}) \left|\int \text{d}\hat{\vec{k}} \frac{g(\hat{\vec{k}})}{\sqrt{G}}\right|^2.
\end{align}
Therefore, we once again retrieve an absorption rate proportional to the absorption cross section, as expected.

\section{Squeezing Terms}\label{sec:squeezing}
Squeezing terms are usually encountered in nonlinear optics. They can arise, for example, from higher order terms of the form $a_{\text{p}}^\dagger a_{\text{s}} a_{\text{s}}$, where subscripts p and s denote the pump and signal modes with $\omega_{\text{p}} = 2\omega_{\text{s}}$. For coherent pumping the operator $a_{\text{p}}$ can be replaced by the classical amplitude of the pump light, which gives rise to a squeezing term of the form $a_{\text{s}}a_{\text{s}}$. In the absence of a pump field $\braket{a_{\text{p}}} = 0$ and, consequently, no squeezing terms appear. This is in contrast to the present theory, where squeezing terms are independent of a classical amplitude and, therefore, independent of any external pumping. As a result, squeezing can occur even in the few photon regime or in the ground state.

In this section, we show that the squeezing terms in Eq. \eqref{eq:complete_hamiltonian} can indeed achieve squeezing of a single cavity mode, but also give rise to thermal-like background radiation. We define 'thermal-like' states here as states showing the same photon bunching behaviour as thermal states obeying Bose-Einstein statistics. However, the ground state of our theory differs from a thermal state due to non-vanishing correlations between different field modes. This is a direct consequence of the Hamiltonian in Eq. \eqref{eq:complete_hamiltonian} including spectral mixing.

\subsection{Scattering Theory}
We start by considering an arbitrary, finite optical structure with absorptive losses. This general case requires perturbative techniques to obtain approximate results. In App. \ref{app:thermal_spectrum} we demonstrate that to leading order in the coupling coefficients $\zeta(\vec{k},\lambda;\vec{k'},\lambda')$ the ground state photon number spectrum reads
\begin{align} \label{eq:thermal_spectrum}
\bra{0}a^\dagger_{\vec{k},\lambda}a_{\vec{k},\lambda}\ket{0} \approx& \sum_{\lambda'}\int \text{d}^3k' \frac{4}{\hbar^2} \frac{\left|\zeta(\vec{k},\lambda;\vec{k'},\lambda')\right|^2}{(\omega+\omega')^2}.
\end{align}
No assumptions about the system geometry and spectrum were made in the derivation of Eq. \eqref{eq:thermal_spectrum}, which is therefore valid for a large class of optical systems. In App.~\ref{app:thermal_spectrum} it is also shown that the modified ground state due to the squeezing terms is indeed thermal-like to leading order in perturbation theory.

\subsection{Cavity Quantum Electrodynamics}
To gain some physical insight into the background photon spectra of realistic systems, we consider now the special case of a single mode optical cavity with Lorentzian spectrum. Specifically, we assume that the electric field inside the cavity illuminated by a normalized plane wave of wavevector $k$ and polarization $\lambda$ reads
\begin{align} \label{eq:gaussian_mode}
\vec{E}_{\vec{k}, \lambda}(\vec{x}) =& \vec{E}_0(\vec{x}) \delta_{\lambda, \lambda_0} \sqrt{\frac{\Gamma}{2\pi}} \frac{g(\hat{\vec{k}})}{\omega-\omega_0+\ii\Gamma/2}.
\end{align}
Please note that in writing Eq. \eqref{eq:gaussian_mode} we assumed polarization selectivity of the mode. From this point on we will only consider the polarization $\lambda_0$ and drop the polarization indices. Further assuming that dispersion is negligible over the spectral width of the cavity mode, the coupling parameter (of the input modes) in Eq. \eqref{eq:thermal_coefficient} becomes
\begin{align} \label{eq:thermal_coefficient_cavity}
\zeta(\vec{k}, \vec{k'}) \approx& \int \text{d}^3x\frac{\ii}{2} \epsilon_0\epsilon''(\omega_0, \vec{x})\vec{E}_0(\vec{x})\cdot\vec{E}_0(\vec{x}) \frac{\Gamma}{2\pi} \frac{g(\hat{\vec{k}})}{\omega-\omega_0+\ii\Gamma/2}\frac{g(\hat{\vec{k'}})}{\omega'-\omega_0+\ii\Gamma/2}\nonumber \\
=& \frac{\ii}{2}\beta\left(\frac{\Gamma}{2}\right)^2 \frac{g(\hat{\vec{k}})}{\omega-\omega_0+\ii\Gamma/2}\frac{g(\hat{\vec{k'}})}{\omega'-\omega_0+\ii\Gamma/2},\\
\label{eq:squeezing_parameter}
\beta :=& \int \text{d}^3x \epsilon_0\epsilon''(\omega_0,\vec{x}) \vec{E}_{\text{max}}(\vec{x})\cdot\vec{E}_{\text{max}}(\vec{x}).
\end{align}
Inserting Eq. \eqref{eq:thermal_coefficient_cavity} into \eqref{eq:thermal_spectrum}, we find the result for the thermal-like photon spectrum of a Lorentzian cavity:
\begin{align} \label{eq:lorentzian_thermal_spectrum}
\bra{0}a^\dagger_{\vec{k},\lambda_0}a_{\vec{k},\lambda_0}\ket{0} =& \frac{\pi G \Gamma^3}{32c_0^3} \frac{\left|\beta\right|^2}{\hbar^2} \frac{\left|g(\hat{\vec{k}})\right|^2}{(\omega-\omega_0)^2+(\Gamma/2)^2} \cdot F(\omega), \\
F(\omega) =& \int_0^\infty d\omega' \frac{4\omega'^2}{(\omega+\omega')^2} \frac{\Gamma}{2\pi} \frac{1}{(\omega'-\omega_0)^2+(\Gamma/2)^2}. \nonumber
\end{align}

Equation \eqref{eq:lorentzian_thermal_spectrum} describes a Lorentzian spectrum modified by a shape factor $F(\omega)$. If the linewdith of the cavity mode is much smaller than the resonance frequency, $\Gamma/2 \ll \omega_0$, the shape factor can be approximated as $F(\omega) \approx 1$ for $\omega \approx \omega_0$. In this case the background photon number spectrum is a Lorentzian, just like the cavity spectrum. Integration of Eq. \eqref{eq:lorentzian_thermal_spectrum} in this case yields the total number of background photons
\begin{align}
N_t \approx& \frac{\pi^2 G^2 \Gamma^2\omega_0^2}{16c_0^6} \frac{\left|\beta\right|^2}{\hbar^2}.
\end{align}

\begin{figure}
\begin{center}
\includegraphics[width=.5\textwidth]{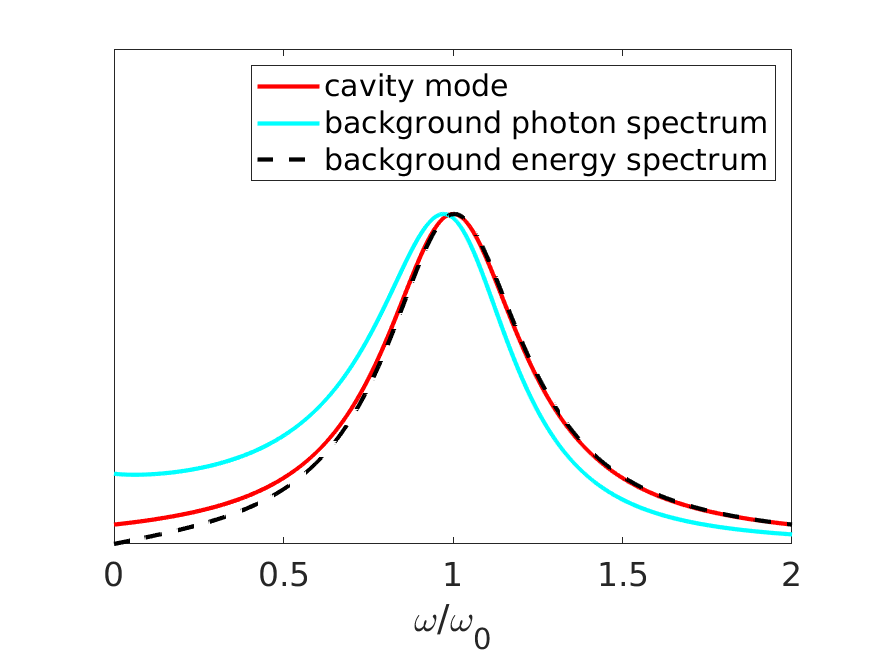}
\caption{Background photon number and energy spectra of a broad Lorentzian resonance.}
\label{fig:thermal_spectrum}
\end{center}
\end{figure}

However, if the linewdith of the cavity mode is comparable to the resonance frequency, the photon number spectrum gets distorted. An example is shown in Fig. \ref{fig:thermal_spectrum}, where $F(\omega)$ was evaluated numerically. While the photon number spectrum preserves its Lorentzian shape, it gets slightly red shifted relative to cavity mode. This is due to the fact that the background photon number spectrum is not identical to the background energy spectrum, which can be obtained from the former by multiplying with the photon energy $\hbar \omega$. The background energy spectrum is also shown in Fig. \ref{fig:thermal_spectrum} and is seen to coincide almost perfectly with the cavity spectrum, except for the region $\omega \approx 0$. This deviation is necessary, since measureable fluctuations can not occur at zero frequency.

We now turn to the dynamics of photons occupying a Lorentzian cavity mode. We use the formalism introduced in \cite{Oppermann2018a}, which requires us to assume that the spectral linewidth is small compared to the resonance frequency $\Gamma/2 << \omega_0$. In the following, we show that our theory predicts ground state squeezing in absorptive optical cavities even in the absence of nonlinearities. Furthermore, we derive an upper bound for the magnitude of quadrature operator squeezing.

We start by deriving the Heisenberg equations of the photon annihilation operators from Eq. \eqref{eq:complete_hamiltonian} while assuming $\kappa(\vec{k}, \vec{k'}) = 0$
\begin{align} \label{eq:scattering_heisenberg_counter}
\dot{a}_{\vec{k}} =& -\ii\omega a_{\vec{k}} + \frac{\ii}{\hbar} \int \text{d}^3k' 2\zeta^*(\vec{k},\vec{k'}) a^\dagger_{\vec{k'}}.
\end{align}
We proceed in accordance with \cite{Oppermann2018a} by introducing the cavity mode operator
\begin{align}
a =& \int \text{d}^3k \frac{c_0^{3/2}}{\sqrt{G}\omega} \sqrt{\frac{\Gamma}{2\pi}} \frac{g(\hat{\vec{k}})}{\omega-\omega_0+\ii\Gamma/2} a_{\vec{k}}, \\
G :=& \int \text{d}\hat{\vec{k}} \left|g(\hat{\vec{k}})\right|^2, \nonumber
\end{align}
and deriving its Heisenberg equation from \eqref{eq:scattering_heisenberg_counter} under the assumption $\Gamma \ll \omega_0$. The result is
\begin{align} \label{eq:cavity_equation_counter}
\dot{a} =& (-\ii\omega_0-\Gamma/2)a - \ii F_0 + B^* a^\dagger, \\
F_0 :=& \int \text{d}^3k \frac{c_0^{3/2}}{\sqrt{G}\omega} \sqrt{\frac{\Gamma}{2\pi}} g^*(\hat{\vec{k}})(\omega-\omega_0)^n a_{\vec{k}}, \nonumber \\
B :=& \frac{G}{4\pi} \frac{2\pi^2\omega_0^2\Gamma}{c_0^3} \frac{\beta}{\hbar}, \nonumber
\end{align}
where $\beta$ is given by Eq. \eqref{eq:squeezing_parameter}.

The ground state squeezing of the system can be calculated by assuming no external pump field. For simplicity we assume that $B \in \mathbb{R}$, which can always be achieved by adding an appropriate constant phase to the electric field. The calculation of the steady state of the cavity mode is perforemd in App. \ref{app:vacuum_squeezing}. The quantities we are interested in are the variances of the quadrature operators
\begin{align} \label{eq:quadrature_definition}
X_{\phi} :=& \frac{1}{2} \left[\ee^{\ii\phi/2}a + \ee^{-\ii\phi/2}a^\dagger\right],
\end{align}
which read
\begin{align} \label{eq:quadrature_variance}
\braket{X_{\phi}^2} - \braket{X_{\phi}}^2 =& \frac{1}{4} + \frac{1}{4}\left[\frac{1+\frac{\cos(\phi)B\Gamma/2}{\omega_0^2+(\Gamma/2)^2}+\frac{\sin(\phi)B\omega_0}{\omega_0^2+(\Gamma/2)^2}}{1-\frac{B^2}{\omega_0^2+(\Gamma/2)^2}} B + \cos(\phi)\frac{\Gamma}{2}+\sin(\phi)\omega_0\right] \frac{B}{\omega_0^2+(\Gamma/2)^2} \nonumber \\
\approx& \frac{1}{4} \left[1 + \frac{B}{\omega_0}\sin(\phi) \right],
\end{align}
We see that maximum squeezing occurs for $\phi = 3\pi/2$, where the variance takes on its minimal value
\begin{align}\label{eq:quadrature_variance_minimum}
\braket{X_{3\pi/2}^2} - \braket{X_{3\pi/2}}^2 \approx \frac{1}{4} \left[1 - \frac{B}{\omega_0} \right].
\end{align}

According to the above discussion, whether or not squeezing in absorptive optical systems is appreciable depends largely on the value of $B/\omega_0$. In the following we derive an upper bound for this parameter. It is interesting to note, that the parameter $\beta$ in Eq. \eqref{eq:squeezing_parameter} bears a formal similiarity to the ohmic loss integral of electrodynamics. To be more precise, from the triangle inequality for integrals we find the following relation between the parameter $\beta$, the energy content of a single photon $\hbar\omega_0$ and the absorptive loss rate $\gamma_{\text{abs}}$:
\begin{align}
\omega_0 \left|\beta\right| \leq& \int \text{d}^3x \omega_0\epsilon_0\epsilon''(\omega_0,\vec{x}) \left|\vec{E}_{\text{max}}(\vec{x})\right|^2 = \frac{1}{2}\cdot\hbar\omega_0\cdot\gamma_{\text{abs}}.
\end{align}
where in the last step we used the fact that the ohmic loss rate is just the energy content of the field times the absorption rate. From the above, it follows that $\beta$ is bounded from above according to
\begin{align} \label{eq:squeezing_bound}
\left|\beta\right| \leq& \hbar\frac{\gamma_{\text{abs}}}{2}.
\end{align}
 Please note that the cavity mode parameter $B$ in Eq. \eqref{eq:cavity_equation_counter} can be written as the parameter $\beta$ times a quantity of dimension $\text{m}^{-3}$. This quantity is the local density of states (LDOS) \cite{Oppermann2018a}
\begin{align}
\text{LDOS} =& \frac{G}{4\pi} \frac{2\pi^2\omega_0^2\Gamma}{c_0^3}.
\end{align}
Since the LDOS connects the quantities $\beta$ and $\gamma_{\text{abs}}$ with the cavity parameter $B = \text{LDOS}\times\beta$ and the cavity mode absorption rate $\Gamma_{\text{abs}} = \text{LDOS}\times\gamma_{\text{abs}}$, Eq. \eqref{eq:squeezing_bound} yields the following relation:
\begin{align} \label{eq:squeezing_constraint}
\left|B\right| \leq& \frac{\Gamma_{\text{abs}}}{2} \ll \omega_0.
\end{align}
But this means that the maximum quadrature squeezing is limited according to 
\begin{align}
\braket{X_{3\pi/2}^2} - \braket{X_{3\pi/2}}^2 \geq \frac{1}{4} \left[1 - \frac{\Gamma_{\text{abs}}/2}{\omega_0} \right].
\end{align}

To summarize: we showed that the counter-rotating terms give rise to a non-trivial ground state, which corresponds to the thermal-like photon spectrum of the absorptive optical system. The existence of fluctuations in absorptive systems is dictated by the fluctuation-dissipation theorem and, therefore, expected. Such fluctuations are absent in optical systems with purely radiative losses, since these do not exhibit dissipation of energy. We further demonstrated that squeezing can arise in absorptive optical cavities even in the absence of nonlinearities and derived a lower bound for the quadrature variance.

\section{Photon Conversion Terms}\label{sec:coupling}
The last term of the Hamiltonian in Eq. \eqref{eq:complete_hamiltonian} describes conversion between the different harmonic oscillator eigenmodes of the system. The phenomenological applications of this off-resonant photon transfer are discussed in the following.

\subsection{Scattering Theory}
From a scattering standpoint off-resonant photon transfer might have two implications. First, an additional scattering mechanism might arise due coherent energy exchange between different modes. This mechanism would modify the results obtained from classical scattering theory and, therefore, Maxwell's equations. Secondly, the photon excitation spectrum might be modified, leading to a modified free space dispersion relation $\widetilde{\omega}(\vec{k}, \lambda) \neq c_0k$. Both effects would, of course, be in stark contradiction to the theory of electromagnetism and its countless experimental verifications. However, both effects can be shown to be non-existent due to conservation of energy and the very specific form of the coupling constant in Eq. \eqref{eq:coupling_coefficient}. The details can be found in App. \ref{app:photon_excitation_spectrum}.

\subsection{Cavity Quantum Electrodynamics}
Let us now turn to a scenario where spatially and spectrally confined modes of light exist in the system. Assume that the optical system features two Lorentzian resonances at frequencies $\omega_{1,2}$ of linewidth $\Gamma_{1,2}$. Assuming that dispersion is neglegible over the intervals $[\omega_i-\Gamma_i/2, \omega_i+\Gamma_i/2]$, the coupling coefficient in Eq. \eqref{eq:coupling_coefficient} becomes
\begin{align}
\kappa(\vec{k}, \vec{k'}) \approx& \ii \chi \left(\frac{\overline{\Gamma}}{2}\right)^2 \frac{g_1^*(\hat{\vec{k}})}{\omega-\omega_1-\ii\Gamma_1/2}\frac{g_2(\hat{\vec{k'}})}{\omega'-\omega_2+\ii\Gamma_2/2} \nonumber \\
&-\ii \chi^* \left(\frac{\overline{\Gamma}}{2}\right)^2 \frac{g_2^*(\hat{\vec{k}})}{\omega-\omega_2-\ii\Gamma_2/2}\frac{g_1(\hat{\vec{k'}})}{\omega'-\omega_1+\ii\Gamma_1/2} , \\
\label{eq:coupling_constant_lorentzian}
\chi :=& \int \text{d}^3x \epsilon_0 \frac{\epsilon''(\omega_1, \vec{x}) - \epsilon''(\omega_2, \vec{x})}{2}\vec{E}^*_{\text{max}, 1} \cdot \vec{E}_{\text{max}, 2},
\end{align}
where we introduced the notation
\begin{align}
\overline{O} := \sqrt{O_1O_2}
\end{align}
Using once again the formalism introduced in \cite{Oppermann2018a}, one can derive a system of Heisenberg equations for the two resonator modes
\begin{align} \label{eq:coupled_oscillator_modes}
\dot{a}_1 \approx& (-\ii\omega_1-\Gamma_1/2)a_1 - \ii  F^{(1)}_0 + X a_2, \nonumber \\
\dot{a}_2 \approx& (-\ii\omega_2-\Gamma_2/2)a_2 - \ii  F^{(2)}_0 - X^* a_1, \\
F^{(i)}_0 :=& \int \text{d}^3k \frac{c_0^{3/2}}{G_i\omega} \sqrt{\frac{\Gamma_i}{2\pi}} g_i^*(\hat{\vec{k}})(\omega-\omega_i)^n a_{\vec{k}}, \nonumber \\
X :=& \frac{\overline{G}}{4\pi}\frac{2\pi^2\overline{\omega}^2\overline{\Gamma}}{c_0}^3 \frac{\chi}{\hbar}.
\end{align}
The system of equations \eqref{eq:coupled_oscillator_modes} has the form of two coupled harmonic resonators.

It is worth noting that, while the coupling constant in Eq. \eqref{eq:coupling_constant_lorentzian} contains the imaginary part of the permittivity, it is possible to couple an absorptionless mode with $\epsilon''(\omega_1) \approx 0$ to an absorptive mode with $\epsilon''(\omega_2) > 0$. This makes the above coupling scheme especially interesting in the context of hybrid plasmonic-photonic systems, where the long lifetime of a high-Q dielectric cavity can be combined with the strong field enhancement of a plasmonic structure. The plasmonic structure to be considered could be, e.g. metallic nanostructures but also graphene based plasmonic antennas.

\section{Conclusion}
In this work we present a method of second quantization for optical systems with absorptive losses. Unlike other methods to describe losses in quantum systems our approach does not require the intoduction of a thermal bath. Instead the dissipation of photons is already encoded in the classical scattering modes used during second quantization. We show that additional terms arising during second quantization do not contradict the experimental reality of absorptive systems and are even able to describe thermal-like photon fluctuations. Our theory leads us to predict a new type of coupling between detuned resonator modes, which relies on the dispersive nature of absorptive optical media.

\section*{Acknowledgements}

J. O. acknowledges support from the Karlsruhe School of Optics and Photonics (KSOP). C.R. acknowledges support from the DFG project No. RO 3640/8-1. We also also wish to thank the Deutscher Akademischer Austauschdienst (PPP Poland) and the Ministry of Science and Higher Education in Poland for support.

\bibliography{references}

\appendix

\section{Derivation of the Hamiltonian}\label{app:hamiltonian_derivation}
We treat the terms of the field Hamiltonian \eqref{eq:quantized_hamiltonian} individually by inserting the expansions \eqref{eq:a_expansion}, \eqref{eq:pia_expansion}, \eqref{eq:pic_expansion} and \eqref{eq:j_expansion} and integrating over space. The momentum term becomes
\begin{align}
\int \text{d}^3x \epsilon_0^{-1} \vec{\Pi}_A(\vec{x})\vec{\Pi}_C(\vec{x}) =& \int \text{d}^3x \epsilon_0^{-1} \vec{\Pi}_C(\vec{x})\vec{\Pi}_A(\vec{x}) \nonumber\\
=& -\frac{1}{4}\sum_{\lambda,\lambda'}\int \text{d}^3k \text{d}^3k' \int \text{d}^3x \epsilon_0\epsilon'(\omega,\vec{x}) \nonumber \\
&\left[\vec{E}_{\vec{k},\lambda}(\vec{x})\vec{E}_{\vec{k'},\lambda'}(\vec{x})a_{\vec{k},\lambda}a_{\vec{k'},\lambda'}-\vec{E}_{\vec{k},\lambda}(\vec{x})\vec{E}^*_{\vec{k'},\lambda'}(\vec{x})a_{\vec{k},\lambda}a^\dagger_{\vec{k'},\lambda'} \right. \nonumber \\
&\left.-\vec{E}^*_{\vec{k},\lambda}(\vec{x})\vec{E}_{\vec{k'},\lambda'}(\vec{x})a^\dagger_{\vec{k},\lambda}a_{\vec{k'},\lambda'}+\vec{E}^*_{\vec{k},\lambda}(\vec{x})\vec{E}^*_{\vec{k'},\lambda'}(\vec{x})a^\dagger_{\vec{k},\lambda}a^\dagger_{\vec{k'},\lambda'}\right] \nonumber \\
=& \frac{1}{8}\sum_{\lambda=1,2}\int \text{d}^3k \hbar\omega \left[a^\dagger_{\vec{k},\lambda}a_{\vec{k},\lambda}+a_{\vec{k},\lambda}a^\dagger_{\vec{k},\lambda}\right] \nonumber \\
&-\frac{1}{4}\sum_{\lambda,\lambda'}\int \text{d}^3k \text{d}^3k' \int \text{d}^3x \epsilon_0\epsilon'(\omega,\vec{x}) \nonumber \\
&\left[\vec{E}_{\vec{k},\lambda}(\vec{x})\vec{E}_{\vec{k'},\lambda'}(\vec{x})a_{\vec{k},\lambda}a_{\vec{k'},\lambda'}+\vec{E}^*_{\vec{k},\lambda}(\vec{x})\vec{E}^*_{\vec{k'},\lambda'}(\vec{x})a^\dagger_{\vec{k},\lambda}a^\dagger_{\vec{k'},\lambda'}\right].
\end{align}
For the evaluation of the curl term we use partial integration and the wave equation \eqref{eq:wave_equation} to obtain
\begin{align}
\int \text{d}^3x \frac{1}{2\mu_0}\left(\nabla\times\vec{A}(\vec{x})\right)^2 =& \int \text{d}^3x \frac{1}{2\mu_0}\vec{A}(\vec{x})\cdot\nabla\times\nabla\times\vec{A}(\vec{x}) \nonumber \\
=& \frac{1}{2}\sum_{\lambda, \lambda'}\int \text{d}^3k \text{d}^3k' \int \text{d}^3x \epsilon_0 \nonumber \\
&\left[\epsilon(\omega', \vec{x})\vec{E}_{\vec{k},\lambda}(\vec{x})\vec{E}_{\vec{k'},\lambda'}(\vec{x})a_{\vec{k},\lambda}a_{\vec{k'},\lambda'}+\epsilon^*(\omega', \vec{x})\vec{E}_{\vec{k},\lambda}(\vec{x})\vec{E}^*_{\vec{k'},\lambda'}(\vec{x})a_{\vec{k},\lambda}a^\dagger_{\vec{k'},\lambda'}\right. \nonumber \\
&+\left.\epsilon(\omega', \vec{x})\vec{E}^*_{\vec{k},\lambda}(\vec{x})\vec{E}_{\vec{k'},\lambda'}(\vec{x})a^\dagger_{\vec{k},\lambda}a_{\vec{k'},\lambda'}+\epsilon^*(\omega', \vec{x})\vec{E}^*_{\vec{k},\lambda}(\vec{x})\vec{E}^*_{\vec{k'},\lambda'}(\vec{x})a^\dagger_{\vec{k},\lambda}a^\dagger_{\vec{k'},\lambda'}\right] \nonumber \\
=&\frac{1}{4}\sum_{\lambda=1,2}\int \text{d}^3k \hbar\omega \left[a^\dagger_{\vec{k},\lambda}a_{\vec{k},\lambda} + a_{\vec{k},\lambda}a^\dagger_{\vec{k},\lambda}\right] \nonumber \\
&+\frac{1}{2}\sum_{\lambda,\lambda'}\int \text{d}^3k \text{d}^3k' \int \text{d}^3x \epsilon_0\epsilon'(\omega,\vec{x}) \nonumber \\
&\left[\vec{E}_{\vec{k},\lambda}(\vec{x})\vec{E}_{\vec{k'},\lambda'}(\vec{x})a_{\vec{k},\lambda}a_{\vec{k'},\lambda'}+\vec{E}^*_{\vec{k},\lambda}(\vec{x})\vec{E}^*_{\vec{k'},\lambda'}(\vec{x})a^\dagger_{\vec{k},\lambda}a^\dagger_{\vec{k'},\lambda'}\right] \nonumber \\
&+\frac{1}{2}\sum_{\lambda,\lambda'}\int \text{d}^3k \text{d}^3k' \int \text{d}^3x \ii\epsilon_0\epsilon''(\omega',\vec{x}) \nonumber \\
&\left[\vec{E}_{\vec{k},\lambda}(\vec{x})\vec{E}_{\vec{k'},\lambda'}(\vec{x})a_{\vec{k},\lambda}a_{\vec{k'},\lambda'}-\vec{E}_{\vec{k},\lambda}(\vec{x})\vec{E}^*_{\vec{k'},\lambda'}(\vec{x})a_{\vec{k},\lambda}a^\dagger_{\vec{k'},\lambda'}\right. \nonumber \\
&\left.+\vec{E}^*_{\vec{k},\lambda}(\vec{x})\vec{E}_{\vec{k'},\lambda'}(\vec{x})a^\dagger_{\vec{k},\lambda}a_{\vec{k'},\lambda'}-\vec{E}^*_{\vec{k},\lambda}(\vec{x})\vec{E}^*_{\vec{k'},\lambda'}(\vec{x})a^\dagger_{\vec{k},\lambda}a^\dagger_{\vec{k'},\lambda'}\right].
\end{align}
Lastly, the ohmic loss term becomes
\begin{align}
-\int \text{d}^3x \frac{1}{2} \vec{A}(\vec{x}) \cdot \vec{j}(\vec{x}) =& -\int \text{d}^3x \frac{1}{2} \vec{j}(\vec{x}) \cdot \vec{A}(\vec{x}) \nonumber \\
=& -\frac{1}{2}\sum_{\lambda,\lambda'} \int \text{d}^3k \text{d}^3k' \int \text{d}^3x \ii\epsilon_0\epsilon''(\omega',\vec{x}) \nonumber \\
&\left[\vec{E}_{\vec{k},\lambda}(\vec{x})\vec{E}_{\vec{k'},\lambda'}(\vec{x})a_{\vec{k},\lambda}a_{\vec{k'},\lambda'}-\vec{E}_{\vec{k},\lambda}(\vec{x})\vec{E}^*_{\vec{k'},\lambda'}(\vec{x})a_{\vec{k},\lambda}a^\dagger_{\vec{k'},\lambda'}\right. \nonumber \\
&\left.+\vec{E}^*_{\vec{k},\lambda}(\vec{x})\vec{E}_{\vec{k'},\lambda'}(\vec{x})a^\dagger_{\vec{k},\lambda}a_{\vec{k'},\lambda'}-\vec{E}^*_{\vec{k},\lambda}(\vec{x})\vec{E}^*_{\vec{k'},\lambda'}(\vec{x})a^\dagger_{\vec{k},\lambda}a^\dagger_{\vec{k'},\lambda'}\right].
\end{align}
Adding up all of the above contributions and ignoring constant terms we finally arrive at
\begin{align}
H =& \sum_{\lambda=1,2}\int \text{d}^3k \hbar\omega a^\dagger_{\vec{k},\lambda}a_{\vec{k},\lambda} \nonumber \\
&- \sum_{\lambda,\lambda'} \int \text{d}^3k \text{d}^3k' \left[\zeta(\vec{k}, \lambda; \vec{k'}, \lambda') a_{\vec{k},\lambda}a_{\vec{k'},\lambda'} + \zeta^*(\vec{k}, \lambda; \vec{k'}, \lambda') a^\dagger_{\vec{k},\lambda}a^\dagger_{\vec{k'},\lambda'}\right] \nonumber \\
&- \sum_{\lambda,\lambda'} \int \text{d}^3k \text{d}^3k' \kappa(\vec{k}, \lambda; \vec{k'}, \lambda') a^\dagger_{\vec{k},\lambda}a_{\vec{k'},\lambda'},
\end{align}
where the functions $\zeta$ and $\kappa$ are defined as
\begin{align}
\zeta(\vec{k}, \lambda; \vec{k'}, \lambda') = \int \text{d}^3x \frac{\ii}{4}\epsilon_0\left[\epsilon''(\omega, \vec{x})+\epsilon''(\omega', \vec{x})\right] \vec{E}_{\vec{k},\lambda}(\vec{x})\cdot\vec{E}_{\vec{k'},\lambda'}(\vec{x}), \\
\kappa(\vec{k}, \lambda; \vec{k'}, \lambda') = \int \text{d}^3x \frac{\ii}{2}\epsilon_0\left[\epsilon''(\omega, \vec{x})-\epsilon''(\omega', \vec{x})\right] \vec{E}^*_{\vec{k},\lambda}(\vec{x})\cdot\vec{E}_{\vec{k'},\lambda'}(\vec{x}).
\end{align}

\section{Free Space Input-Output Relation}\label{app:input_output_relation}
In order to find the operator input-output relations, we need to expand the classical input modes in the classical output modes. This is possible because both sets of modes form complete sets, see \cite{Oppermann2018b}. The input modes are of the form
\begin{align} \label{eq:input_mode}
\vec{E}_{\vec{k},\lambda}(\vec{x}) = \vec{E}^{(\vec{k}, \lambda)}_0 \ee^{\ii\vec{k}\vec{x}} + \vec{E}^{(\vec{k}, \lambda)}_S(\vec{x}),
\end{align}
where the first term corresponds to the incident plane wave and the second term to the scattered field. The calculations are significantly simplified, if one uses the fact that any solution to Maxwell's equations is completely determined by its far field. In \cite{Oppermann2018b} it is shown that for the purpose of normalization integrals, near field contributions can be ignored. This is because near field corrections are finite and, therefore, lead to vanishing contributions compared to terms containing $\delta$-distributions when taking the physical plane wave limit. Consequently, for the purpose of calculating modal expansion coefficients, we can replace the scattered field by its far field form
\begin{align}
\vec{E}^{(\vec{k},\lambda)}_S(\vec{x}) \approx \altvec{A}^{(\vec{k},\lambda)}(\hat{\vec{x}}) \frac{\ee^{\ii k r}}{kr},
\end{align}
which only contains outgoing contributions. We now define an inverse-scattering term, which only consists of incoming contributions
\begin{align}
\vec{E}^{(\vec{k},\lambda)}_{IS}(\vec{x}) \approx \altvec{A}^{(\vec{k},\lambda)}(-\hat{\vec{x}}) \frac{\ee^{-\ii k r}}{kr},
\end{align}
and rewrite the input mode in Eq. \eqref{eq:input_mode} as
\begin{align} \label{eq:input_mode_rewritten}
\vec{E}_{\vec{k},\lambda}(\vec{x}) = \vec{E}^{(\vec{k}, \lambda)}_0 \ee^{\ii\vec{k}\vec{x}} + \vec{E}^{(\vec{k}, \lambda)}_{IS}(\vec{x}) + \vec{E}^{(\vec{k}, \lambda)}_S(\vec{x}) - \vec{E}^{(\vec{k}, \lambda)}_{IS}(\vec{x}).
\end{align}
It is now a straightforward matter to show that the last two terms of Eq. \eqref{eq:input_mode_rewritten} can be expanded in plane waves, by simply performing the Fourier transform
\begin{align}
\frac{1}{(2\pi)^2} \int \text{d}^3x \ee^{-\ii\vec{k'}\vec{x}}\left[\vec{E}^{(\vec{k}, \lambda)}_S(\vec{x}) - \vec{E}^{(\vec{k}, \lambda)}_{IS}(\vec{x})\right] = \frac{\ii}{2\pi k k'} \altvec{A}^{(\vec{k},\lambda)}(\hat{\vec{k'}}) \delta(k-k').
\end{align}
Applying the inverse Fourier transform now leads to
\begin{align} \label{eq:expanded_scattering_amplitude}
\vec{E}^{(\vec{k}, \lambda)}_S(\vec{x}) - \vec{E}^{(\vec{k}, \lambda)}_{IS}(\vec{x}) =& \frac{\ii}{2\pi} \int_{k=k'} d\hat{\vec{k'}} \altvec{A}^{(\vec{k},\lambda)}(\hat{\vec{k'}}) \ee^{\ii\vec{k'}\vec{x}} \nonumber \\
=& \frac{\ii}{2\pi} \sum_{\lambda=1,2}\int_{k=k'} d\hat{\vec{k'}} \frac{\altvec{A}^{(\vec{k},\lambda)}(\hat{\vec{k'}})\cdot\vec{E}^{*(\vec{k'}, \lambda')}_0}{\left|\vec{E}^{(\vec{k'}, \lambda')}_0\right|^2} \vec{E}^{(\vec{k'}, \lambda')}_0 \ee^{\ii\vec{k'}\vec{x}},
\end{align}
where in the last step the two non-zero vector components of the scattering amplitude where projected onto the two different polarization directions of the incident fields. Inserting Eq. \eqref{eq:expanded_scattering_amplitude} into Eq. \eqref{eq:input_mode_rewritten} gives us an expression that consists solely of plane wave and incoming wave contributions. Since an output mode's outgoing contributions are completely contained in its plane wave part, the relation between input and output modes can be found by simply matching the plane wave components of Eq. \eqref{eq:input_mode_rewritten} to the plane wave components of the outgoing modes, leading to
\begin{align}
\vec{E}^{\text{(in)}}_{\vec{k},\lambda}(\vec{x}) = \vec{E}^{\text{(out)}}_{\vec{k}, \lambda}(\vec{x}) + \frac{\ii}{2\pi} \sum_{\lambda=1,2}\int_{k=k'} d\hat{\vec{k'}} \frac{\altvec{A}^{\vec{k},\lambda}(\hat{\vec{k'}})\cdot\vec{E}^{*(\vec{k'}, \lambda')}_0}{\left|\vec{E}^{(\vec{k'}, \lambda')}_0\right|^2} \vec{E}^{\text{(out)}}_{\vec{k'}, \lambda'}(\vec{x}).
\end{align}
All that is left to do is rewriting the positive frequency part of the electric field operator to obtain
\begin{align}
\vec{E}^{(+)}(\vec{x}) =& \sum_{\lambda}\int \text{d}^3k \vec{E}^{\text{(in)}}_{\vec{k},\lambda}(\vec{x})a^{\text{(in)}}_{\vec{k},\lambda} \nonumber \\
=& \sum_{\lambda}\int \text{d}^3k \left[\vec{E}^{\text{(out)}}_{\vec{k}, \lambda}(\vec{x}) + \frac{\ii}{2\pi} \sum_{\lambda'=1,2}\int_{k=k'} d\hat{\vec{k'}} \frac{\altvec{A}^{\vec{k},\lambda}(\hat{\vec{k'}})\cdot\vec{E}^{*(\vec{k'}, \lambda')}_0}{\left|\vec{E}^{(\vec{k'}, \lambda')}_0\right|^2} \vec{E}^{\text{(out)}}_{\vec{k'}, \lambda'}(\vec{x})\right] a^{\text{(in)}}_{\vec{k}, \lambda} \nonumber \\
=& \sum_{\lambda}\int \text{d}^3k \vec{E}^{\text{(out)}}_{\vec{k}, \lambda}(\vec{x}) \left[a_{\vec{k}, \lambda}^{\text{(in)}} + \frac{\ii}{2\pi} \sum_{\lambda'=1,2}\int_{k=k'} d\hat{\vec{k'}} \frac{\altvec{A}^{\vec{k'},\lambda'}(\hat{\vec{k}})\cdot\vec{E}^{*(\vec{k}, \lambda)}_0}{\left|\vec{E}^{(\vec{k}, \lambda)}_0\right|^2} a^{\text{(in)}}_{\vec{k'}, \lambda'}\right],
\end{align}
from which it immediately follows that
\begin{align}
a^{\text{(out)}}_{\vec{k}, \lambda} = a^{\text{(in)}}_{\vec{k}, \lambda} + \frac{\ii}{2\pi} \sum_{\lambda'=1,2}\int_{k=k'} d\hat{\vec{k'}} \frac{\altvec{A}^{\vec{k'},\lambda'}(\hat{\vec{k}})\cdot\vec{E}^{*(\vec{k}, \lambda)}_0}{\left|\vec{E}^{(\vec{k}, \lambda)}_0\right|^2} a^{\text{(in)}}_{\vec{k'}, \lambda'}.
\end{align}

\section{Vacuum Squeezing}\label{app:vacuum_squeezing}
Assume that the system is prepared in the zero photon state $\ket{0}$ at some initial time $t_0$. We will show that in the limit $t \rightarrow \infty$ the system approaches a ground state exhibiting quadrature squeezing. Using the analytical solution for $F_0(t)$ presented in \cite{Oppermann2018a}, it is easily seen that $F_0(t)\ket{0} = 0$ for all times $t$. Therefore, Eq. \eqref{eq:cavity_equation_counter} leads to the following set of expectation value equations
\begin{align}
\frac{\text{d}}{\text{d}t}\braket{a} =& \left(-\ii\omega_0-\Gamma/2\right)\braket{a} + B\braket{a}^*, \\
\frac{\text{d}}{\text{d}t}\braket{a^\dagger a} =& -\Gamma\braket{a^\dagger a} + B \left[\braket{aa} + \braket{aa}^*\right], \\
\frac{\text{d}}{\text{d}t}\braket{aa} =& \left(-2\ii\omega_0-\Gamma\right)\braket{aa} + 2B\braket{a^\dagger a} + B,
\end{align}
where $B\in\mathbb{R}$, see the discussion above Eq. \eqref{eq:quadrature_definition}. The steady-state solution reads
\begin{align}
\braket{a} =& 0, \\
\braket{a^\dagger a} =& \frac{1}{2}\frac{B^2}{\omega_0^2+(\Gamma/2)^2}\frac{1}{1-B^2\left[\omega_0^2+(\Gamma/2)^2\right]^{-2}}, \\
\braket{aa} =& \frac{B}{\ii\omega_0+\Gamma/2}\braket{a^\dagger a} - \frac{1}{2}\frac{1}{\ii\omega_0+\Gamma/2}.
\end{align}
Using the above solution and Eq. \eqref{eq:quadrature_definition} it is straightforward to calculate the quadrature variance
\begin{align}
\braket{X_\phi^2} - \braket{X_\phi}^2 =& \frac{1}{4} + \frac{1}{4} \left(2\braket{a^\dagger a} + \ee^{\ii\phi} \braket{aa} + \ee^{-\ii\phi} \braket{aa}^*\right) \\
=& \frac{1}{4} + \frac{1}{4}\left[\frac{1+\frac{\cos(\phi)B\Gamma/2}{\omega_0^2+(\Gamma/2)^2}+\frac{\sin(\phi)B\omega_0}{\omega_0^2+(\Gamma/2)^2}}{1-\frac{B^2}{\omega_0^2+(\Gamma/2)^2}} B + \cos(\phi)\frac{\Gamma}{2} + \sin(\phi)\omega_0\right] \frac{B}{\omega_0^2+(\Gamma/2)^2},
\end{align}
which is Eq. \eqref{eq:quadrature_variance} of the main text.

\section{Background Photon Spectrum}\label{app:thermal_spectrum}
We want to find the ground state of the Hamiltonian
\begin{align}
H =& \sum_{\lambda=1,2}\int \text{d}^3k \hbar\omega a^\dagger_{\vec{k},\lambda}a_{\vec{k},\lambda} \nonumber \\
&- \sum_{\lambda,\lambda'} \int \text{d}^3k \text{d}^3k' \left[\zeta(\vec{k}, \lambda; \vec{k'}, \lambda') a_{\vec{k},\lambda}a_{\vec{k'},\lambda'} + \zeta^*(\vec{k}, \lambda; \vec{k'}, \lambda') a^\dagger_{\vec{k},\lambda}a^\dagger_{\vec{k'},\lambda'}\right],
\end{align}
which results from Eq. \eqref{eq:complete_hamiltonian} when the coupling terms are dropped. To this end we assume that the counter-rotating terms are adiabatically switched on and follow the temporal evolution of the annihilation operators. The Heisenberg equations of motion read
\begin{align}
\frac{\text{d}}{\text{d}t} a_{\vec{k},\lambda} =& -\ii\omega a_{\vec{k},\lambda} + \frac{\ii}{\hbar}\sum_{\lambda'}\int \text{d}^3k' 2\zeta^*(\vec{k}, \lambda; \vec{k'}, \lambda')\ee^{0^+t}a^\dagger_{\vec{k'}, \lambda'},
\end{align}
which can be formally solved to yield
\begin{align} \label{eq:formal_solution_counter}
a_{\vec{k}, \lambda}(t) =& \ee^{-\ii\omega t} a_{\vec{k}, \lambda}(-\infty) + \frac{\ii}{\hbar}\int_{-\infty}^tdt' \sum_{\lambda'}\int \text{d}^3k' 2\zeta^*(\vec{k}, \lambda; \vec{k'}, \lambda')\ee^{-\ii\omega(t-t')+0^+t'} a^\dagger_{\vec{k'}, \lambda'}(t').
\end{align}
Setting $t=0$ and using the Born approximation allows us to simplify Eq. \eqref{eq:formal_solution_counter} to
\begin{align} \label{eq:thermal_born_solution}
a_{\vec{k}, \lambda}(0) =& a_{\vec{k}, \lambda}(-\infty) + \frac{\ii}{\hbar}\sum_{\lambda'}\int \text{d}^3k' a^\dagger_{\vec{k'}, \lambda'}(-\infty)\int_{-\infty}^0dt'  2\zeta^*(\vec{k}, \lambda; \vec{k'}, \lambda')\ee^{[\ii(\omega+\omega')+0^+]t'}  \nonumber \\
=& a_{\vec{k}, \lambda}(-\infty) + \frac{1}{\hbar}\sum_{\lambda'}\int \text{d}^3k' \frac{2\zeta^*(\vec{k}, \lambda; \vec{k'}, \lambda')}{\omega+\omega'}a^\dagger_{\vec{k'}, \lambda'}(-\infty).
\end{align}
The ground state photon spectrum is now easily obtained by calculating the photon number expectation values
\begin{align}
\braket{0|a^\dagger_{\vec{k}, \lambda}(0)a_{\vec{k}, \lambda}(0)|0} =& \sum_{\lambda'}\int \text{d}^3k' \left|\frac{2}{\hbar}\frac{\zeta(\vec{k}, \lambda; \vec{k'}, \lambda')}{\omega+\omega'}\right|^2.
\end{align}
From Eq. \eqref{eq:thermal_born_solution} we can also calculate the time-delayed second order photon auto-correlation function to find
\begin{align}
g^{(2)}_{\vec{k}, \lambda}(\tau) :=& \frac{\braket{0|a_{\vec{k}, \lambda}^\dagger(\tau)a_{\vec{k}, \lambda}^\dagger(0)a_{\vec{k}, \lambda}(0)a_{\vec{k}, \lambda}(\tau)|0}}{\braket{0|a_{\vec{k}, \lambda}^\dagger(0)a_{\vec{k}, \lambda}(0)|0}^2}\nonumber\\ 
=& 1 + \left|\frac{\sum_{\lambda'}\int \text{d}^3k' \left|\frac{\zeta(\vec{k}, \lambda; \vec{k'}, \lambda')}{\omega+\omega'}\right|^2 \ee^{-\ii \omega' \tau}}{\sum_{\lambda'}\int \text{d}^3k' \left|\frac{\zeta(\vec{k}, \lambda; \vec{k'}, \lambda')}{\omega+\omega'}\right|^2}\right|^2.
\end{align}
We see that $g^{(2)}_{\vec{k}, \lambda}(\tau)$ obeys the identities
\begin{align}
g^{(2)}_{\vec{k}, \lambda}(0) &= 2, \nonumber \\
\lim_{\tau\to\infty}g^{(2)}_{\vec{k}, \lambda}(\tau) &= 1, \nonumber
\end{align}
which shows that the ground state of the system is occupied by thermal-like photons.

\section{Photon Excitation Spectrum}\label{app:photon_excitation_spectrum}
We want to find the excitation spectrum of the Hamiltonian
\begin{align} \label{eq:reduced_hamiltonian_coupling}
H =& \sum_{\lambda=1,2}\int \text{d}^3k \hbar\omega a^\dagger_{\vec{k},\lambda}a_{\vec{k},\lambda} \nonumber \\
&- \sum_{\lambda,\lambda'} \int \text{d}^3k \text{d}^3k' \kappa(\vec{k}, \lambda; \vec{k'}, \lambda') a^\dagger_{\vec{k},\lambda}a_{\vec{k'},\lambda'},
\end{align}
which results from Eq. \eqref{eq:complete_hamiltonian} when neglecting the counter-propagating terms. Since the Hamiltonian \eqref{eq:reduced_hamiltonian_coupling} describes an infinite set of coupled harmonic oscillators, it is theoretically possible to find annihilation operators $\widetilde{a}_{\vec{k},\lambda}$ associated with the dressed modes of the system. Hence, in order to find the resulting spectrum it is enough to consider the single-excitation states of the system, since only integer multiples of their energies can occur in the spectrum. We assume that the interaction term is being adiabatically switched on and track the temporal evolution of single-photon states to find the dressed single-excitation states. Denoting the unitary time-evolution operator with $U(t_{\text{end}}, t_{\text{start}})$ one finds in the interaction picture
\begin{align} \label{eq:time_evolution_operator}
U(0, -\infty)\ket{1_{\vec{k},\lambda}} =& \text{T}\exp\left(\ii \int_{-\infty}^0dt \ee^{\ii(\omega_1-\omega_2-\ii 0^+)t}\sum_{\lambda, \lambda'}\int \text{d}^3k\text{d}^3k'\kappa(\vec{k},\lambda;\vec{k'},\lambda')a^\dagger_{\vec{k},\lambda}a_{\vec{k'},\lambda'}\right)\ket{1_{\vec{k},\lambda}}.
\end{align}
Restricting the problem to the one-photon subspace, Eq. \eqref{eq:time_evolution_operator} becomes
\begin{align} \label{eq:dressed_states}
&U(0, -\infty)\ket{1_{\vec{k},\lambda}} \nonumber \\
&= \text{T}\exp\left(\ii \int_{-\infty}^0dt \sum_{\lambda, \lambda'}\int \text{d}^3k\text{d}^3k' \ee^{\ii(\omega-\omega'-\ii 0^+)t}\kappa(\vec{k},\lambda;\vec{k'},\lambda')\ket{\vec{k},\lambda}\bra{\vec{k'},\lambda'}\right)\ket{1_{\vec{k},\lambda}} \nonumber \\
&= \ket{1_{\vec{k},\lambda}}+\sum_{n=1}^\infty \sum_{\lambda_1,\ldots,\lambda_{n}}\int \text{d}^3k_1 \ldots \int \text{d}^3k_{n}\frac{\kappa(\vec{k}_1,\lambda_1;\vec{k}_2,\lambda_2)}{\omega_1-\omega-\ii0^+}\ldots\frac{\kappa(\vec{k}_n,\lambda_n;\vec{k},\lambda)}{\omega_n-\omega-\ii0^+} \ket{1_{\vec{k}_1,\lambda_1}} \nonumber \\
&= \ket{1_{\vec{k},\lambda}}+\sum_{n=1}^\infty \left(\frac{1}{c_0^3}\right)^n \sum_{\lambda_1,\ldots,\lambda_{n}}\int \text{d}\hat{\vec{k}}_1\int \text{d}\omega_1 \omega_1^2 \ldots \int \text{d}\hat{\vec{k}}_n\int \text{d}\omega_n \omega_n^2 \nonumber \\
& \hspace{8cm} \frac{\kappa(\vec{k}_1,\lambda_1;\vec{k}_2,\lambda_2)}{\omega_1-\omega-\ii0^+}\ldots\frac{\kappa(\vec{k}_n,\lambda_n;\vec{k},\lambda)}{\omega_n-\omega-\ii0^+} \ket{1_{\vec{k}_1,\lambda_1}} \nonumber \\
&= \ket{1_{\vec{k},\lambda}}+\sum_{n=1}^\infty \left(\frac{2\pi\ii\omega^2}{c_0^3}\right)^n \sum_{\lambda_1,\ldots,\lambda_{n}}\int_{k_1 = k} d\hat{\vec{k}}_1 \ldots \int_{k_n = k} d\hat{\vec{k}}_n \nonumber\\
& \hspace{7.5cm} \kappa(\vec{k}_1,\lambda_1;\vec{k}_2,\lambda_2)\ldots\kappa(\vec{k}_n,\lambda_n;\vec{k},\lambda) \ket{1_{\vec{k}_1,\lambda_1}}.
\end{align}
From Eq. \eqref{eq:dressed_states} it becomes clear that only photon states of equal frequency can mix to create the dressed states. However, Eq. \eqref{eq:coupling_coefficient} implies that $\kappa(\vec{k},\lambda;\vec{k'},\lambda') = 0$ if $k=k'$. Therefore, Eq. \eqref{eq:dressed_states} tells us that the single-excitation states of the system are not modified by the presence of the coupling terms. In particular, no modification to the scattering response or dispersion relation occurs.

\end{document}